\documentclass[hyper]{JHEP} 
\usepackage{epsfig}

\newcommand\fverb{\setbox\pippobox=\hbox\bgroup\verb}

\newcommand\fverbdo{\egroup\medskip\noindent%

            \fbox{\unhbox\pippobox}\ }

\newcommand\fverbit{\egroup\item[\fbox{\unhbox\pippobox}]}

\newbox\pippobox

\title{Current Algebra of
the Pure Spinor Superstring in $AdS_5\times S_5$ }

\author{ M. Bianchi and   J. Kluso\v{n}
 \footnote{On leave from Masaryk University, Brno}\\
Dipartimento di Fisica \& Sezione I.N.F.N.\\
Universit\`a di Roma
``Tor Vergata'' \\
Via della Ricerca Scientifica, 1 00133  Roma   ITALY\\
E-mail: \email{Josef.Kluson@roma2.infn.it}}

\preprint{\hepth{0606188}\\
ROM2F/2006/12}

\abstract{We perform a Hamiltonian analysis of the classical type
IIB superstring on $AdS_5 \times S^5$ in the pure spinor approach.
Taking the spatial components of the left-invariant
(super)currents and the pure spinor ghosts as canonical variables,
we compute the classical graded Poisson brackets of the currents
and ghosts and identify the first class constraints associated to
the local $SO(4,1)\times SO(5)$ symmetry and the pure spinor
condition. We then study the properties of the BRST generators and
the Hamiltonian along the constraints. For a natural choice of the
the Lagrange multipliers, we show equivalence of the canonical
equations of motion with the covariant ones. Finally we briefly
discuss the (non) local currents, including the ghost
contribution, that generate the global $PSU(2,2|4)$ symmetry and
its Yangian extension in the present framework.}


\def\be{\begin{equation}}
\def\ee{\end{equation}}
\def\bea{\begin{eqnarray}}
\def\eea{\end{eqnarray}}
\def\be{\begin{equation}}
\def\ee{\end{equation}}
\def\nn{\nonumber}

\def\pa{\partial}

\def\tr{{\mathrm{Tr}}}
\def\str{{\mathrm{Str}}}

\def\pb  #1{\left\{#1\right\}}

\newcommand{\hPhi}{\hat{\Phi}}
\newcommand{\mH}{{\mathcal{H}}}

\newcommand{\halpha}{\hat{\alpha}}
\newcommand{\hbeta}{\hat{\beta}}
\newcommand{\hdelta}{\hat{\delta}}
\newcommand{\hgamma}{\hat{\gamma}}
\newcommand{\hlambda}{\hat{\lambda}}
\newcommand{\hGamma}{\hat{\Gamma}}
\newcommand{\hw}{\hat{w}}
\newcommand{\hN}{\hat{N}}

\newcommand{\com}[1]{\left[#1\right]}

\newcommand{\mL}{{\mathcal{L}}}

\newcommand{\tP}{\tilde{\cal P}}
\newcommand{\cP}{{\cal P}}
\newcommand{\hP}{\hat{\cal P}}
\newcommand{\uc}{\underline{c}}
\newcommand{\ud}{\underline{d}}
\newcommand{\ue}{\underline{e}}
\newcommand{\uf}{\underline{f}}
\newcommand{\ua}{\underline{a}}
\newcommand{\ub}{\underline{b}}
\newcommand{\ug}{\underline{g}}
\newcommand{\uh}{\underline{h}}
\newcommand{\hpi}{\hat{\pi}}

\begin{document}

\section{Introduction}
\label{Intro}

Quantization of the type IIB superstring on $AdS_5\times S_5$
remains an open challenging problem. Some progress has been
achieved through the pure spinor formalism proposed by Berkovits
\cite{Berkovits:1996bf,Berkovits:2000fe,
Berkovits:2000ph,Berkovits:2000nn, Berkovits:2001us} \footnote{For
review of pure spinor formalism in superstring theory, see
\cite{Berkovits:2002zk,Grassi:2005av,Grassi:2003cm,Grassi:2002sr,
Nekrasov:2005wg}.}. In a recent paper \cite{Berkovits:2004xu},
quantum consistency was argued by means of algebraic
renormalization arguments. Vertex operators for massless
excitations have been proposed some time ago
\cite{Berkovits:2000yr} and checked to be classically BRST
invariant \cite{Kluson:2006wq}. However, differently to what
happens in flat spacetime \cite{Berkovits:2002qx}, very little or
nothing is known so far about the emission vertices of massive
states. This is a sad state of affairs, in view of the holographic
correspondence \cite{Maldacena:1997re,Witten:1998qj,Gubser:1998bc}
and in particular of the remarkable agreement found in
\cite{Bianchi:2003wx, Beisert:2003te, Beisert:2004di}
\footnote{For review, see \cite{Bianchi:2004xi, Bianchi:2006gk}.}
 between the
spectrum of single-trace gauge invariant operators in free ${\cal
N} = 4$ SYM and the spectrum of the type IIB superstring on
$AdS_5\times S_5$ extrapolated to the point of higher-spin
symmetry enhancement. As always in physics, the situation should
improve by further exploiting the symmetries of the background.
Because of the presence of the RR 5-form flux, worldsheet currents
are not chirally split as for instance in WZW models. The study of
their quantum OPE may not forgo a classical analysis, which
presents some subtleties in view of the non-trivial role of the
pure spinor ghosts. For this reason, in the present paper, we
study the classical algebra encoded in the (graded) Poisson
brackets of the left-invariant (super)currents $J_{\mu}^A =
Str(g^{-1}\partial_\mu g T^A)$ and the ghost currents. To this end
we resort to a slightly unconventional approach
\cite{Faddeev:1987ph, Korotkin:1997fi} whereby the spatial
components of the (super)currents $J_1^A$, rather than the
supercoset representative itself $g\in PSU(2,2|4)/SO(4,1)\times
SO(5)$, are taken as canonical variables. Along the way, we
identify the first class constraints generating the local
$SO(4,1)\times SO(5)$ symmetry and the gauge transformations
arising from the pure spinor constraints. We explicitly determine
the action of the classical BRST generators on the fundamental
worldsheet fields and currents. We then show that the BRST
generators commute with the Hamiltonian and we also prove that
these BRST generators are nihilpotent along the constraints. A
similar analysis in the more standard approach with $g$ as
canonical variable, has been performed by Chandia for the
heterotic string in the pure spinor formulation
\cite{Chandia:2006ix}.

The plan of the paper is as follows. In Section \ref{BasAdS} we
recall some basic facts about the pure spinor formulation of the
type IIB superstring on $AdS_5\times S_5$. In Section
\ref{Hamanalys}, after identifying the momenta $\Pi_A$ conjugate
to the spatial components of the left-invariant currents $J_1^A$,
we compute the classical graded Poisson brackets of the currents
in a Hamiltonian approach. In section \ref{BRSTHamil}, we study
the BRST generators and  the Hamiltonian of the theory. In section
\ref{Caneqs}, we derive  the canonical equations of motion and
show they are equivalent to the covariant ones for a natural
choice of the Lagrange multipliers. Conservation and nihilpotency
of the BRST charge along the constraints are shown in section
\ref{Consnihil}. In section \ref{Globcurrint} we briefly address
the issue of global symmetries and integrability in the classical
Hamiltonian approach. Section \ref{Conclus} contains our
conclusions and indicates perspectives for future work. Finally
there are two appendices. The first collects our notation and some
important features of $PSU(2,2|4)$. The second describes an
elementary application of the canonical approach presently
exploited to the simple case of a free massless boson.

\section{Pure spinor superstring
in $AdS_5\times S_5$}\label{BasAdS}

As shown in \cite{Berkovits:2000yr,
Berkovits:2004xu,Kluson:2006wq} the  classical action for the
manifestly covariant superstring on $AdS_5\times S_5$ takes the
form
\begin{eqnarray}\label{Minaction}
S = &&-\int d^2x\sqrt{-\eta}\str\left[\frac{1}{2}
\eta^{\mu\nu}\left( J_\mu^{(2)} J_{\nu}^{(2)}+J_\mu^{(1)}J_\nu^{(3)}
+J_\mu^{(3)}J_\nu^{(1)}\right)+ \right.\nonumber\\
&&\left.+ \frac{\epsilon^{\mu\nu}}{4} \left(J^{(1)}_\mu J^{(3)}_\nu-
J^{(3)}_\mu J^{(1)}_\nu\right) \right]
\nonumber \\
&&-\int d^2x\sqrt{-\eta} \str(w_{\mu}\cP^{\mu\nu}
\partial_\nu \lambda
+\hw_{\mu}\tP^{\mu\nu}
\partial_\nu\hlambda
\nonumber \\
&&+N_{\mu} \cP^{\mu\nu}J^{(0)}_\nu+ \hN_{\mu}
\tP^{\mu\nu}J^{(0)}_\nu - \frac{1}{2}N_{\mu}\cP^{\mu\nu}\hN_\nu
-\frac{1}{2}\hN_{\mu}\tP^{\mu\nu} N_\nu) \ ,
\end{eqnarray}
where we have omitted an overall factor \footnote{We work in units
$2\pi\alpha'=1$.} $\sqrt{\lambda}/2\pi = \sqrt{g_s N/\pi}$. We
have assumed the  world-sheet to be a flat two dimensional
space-time with the metric  $\eta=\mathrm{diag}(-1,1)$ and
labeled the world-sheet coordinates as $x^\mu$, with
$\mu,\nu=0,1$. However we also
 use the notation
$x^0=t$, $ x^1=x$ and  $d^2x=dxdt$. We have also introduced the
(chiral) `projectors'
\begin{equation}
\cP^{\mu\nu}= \eta^{\mu\nu}-\epsilon^{\mu\nu}
 \ , \qquad  \tP^{\mu\nu}=
\eta^{\mu\nu}+\epsilon^{\mu\nu}  \ , \qquad \epsilon^{\mu\nu}=
\frac{\varepsilon^{\mu\nu}}{\sqrt{-\eta}} \ ,
\end{equation}
with $\varepsilon^{01}=-\varepsilon^{10}=1$ and
 the left-invariant (super)currents and ghost fields
\begin{eqnarray}\label{matnot}
&&J^{(0)}_\mu=(g^{-1}\partial_\mu g)^{[\uc\ud]}T_{[\uc\ud]} \ ,
\quad J^{(1)}_\mu=(g^{-1}\partial_\mu g)^{\alpha}T_{\alpha} \ ,\nonumber \\
&&J^{(2)}_\mu=(g^{-1}\partial_\mu g)^{\uc}T_{\uc} \ , \quad
J^{(3)}_\mu=(g^{-1}\partial_\mu g)^{\halpha}
T_{\halpha} \ , \nonumber \\
&&\lambda=\lambda^\alpha T_\alpha \ , \quad w_\mu=w_{\mu\alpha}
K^{\alpha\hbeta} T_{\hbeta}  \ , \quad \hlambda=\hlambda^{\halpha}
T_{\halpha}  \ , \quad \hw_\mu=\hw_{\mu\halpha} K^{\halpha\alpha}
T_{\alpha}
 \ , \nonumber \\
&&N_\mu=-\pb{w_\mu,\lambda}=-w_{\mu\beta}\lambda^\alpha
\pb{T_{\hbeta},T_{\alpha}}K^{\beta\hbeta} =
-w_{\mu\beta}K^{\beta\hbeta}
f_{\hbeta\alpha}^{[\uc\ud]}\lambda^\alpha T_{[\uc\ud]} \ ,
\nonumber \\
&&\hat{N}_\mu=-\pb{\hw_\mu,\hlambda}=
-\hw_{\mu\halpha}\hlambda^{\hbeta}
\pb{T_{\hbeta},T_{\alpha}}K^{\halpha\alpha}=
-\hw_{\mu\halpha}K^{\halpha\alpha} f_{\alpha\hbeta}^{[\uc\ud]}
\hlambda^{\hbeta}T_{[\uc\ud]} \  ,
\end{eqnarray}
where $T_A$ are the (super)generators of $psu(2,2|4)$, some of
whose properties can be found in Appendix A, where we define our
notation, and $K^{AB}$ denotes the inverse of the Cartan-Killing
metric.

Following Berkovits, the ghost variables $\lambda$ and
$\hat{\lambda}$ are chosen to satisfy the pure spinor constraints
\begin{equation}\label{purespinri}
\lambda^\gamma \gamma_{\gamma\beta}^{\uc}\lambda^{\beta}=0 \ ,
\quad  \hlambda^{\hgamma}\gamma_{\hgamma \hbeta}^{\uc}
\hlambda^{\hbeta} =0 \ .
\end{equation}
These constraints imply invariance of the action under the gauge
transformations
\begin{eqnarray}\label{locw}
&&\delta w_{\mu\alpha}\cP^{\mu 0}
=-\Lambda_{\uc}(\gamma^{\uc})_\alpha \ , \quad \delta
w_{\mu\alpha}\cP^{\mu 1}= -\Lambda_{\uc}(\gamma^{\uc})_\alpha \  ,
\nonumber \\
&&\delta \hw_{\mu\halpha}\hP^{\mu 0}=
-\hat{\Lambda}_{\uc}(\gamma^{\uc})_{\halpha} \ , \quad \delta
\hw_{\mu\halpha}\hP^{\mu 1}=
\hat{\Lambda}_{\uc}(\gamma^{\uc})_{\halpha} \ .
\end{eqnarray}
Although a promising and thus far consistent formulation of
superstring theories the origin of the pure spinor approach is not
fully understood. Moreover interpreting the pure spinor constraint
(\ref{purespinri}) as the generator of local gauge transformations
(\ref{locw}) involving ($w_\mu,\hw_\mu$) suggests that this
symmetry should be gauge fixed at the quantum level in some way.
There are many proposals as how to deal with the pure spinor
constraint \cite{Berkovits:2005bt,Grassi:2003kq,Aisaka:2005vn,
Gaona:2005yw,Oda:2005sd,Chesterman:2002ey} with no definite widely
accepted conclusion.


\section{Hamiltonian analysis}
\label{Hamanalys}

In this section we are going to  perform the Hamiltonian analysis
of the  action (\ref{Minaction}). Our analysis is based on  the
approach introduced in \cite{Faddeev:1987ph} and recently used in
the context of the GS superstring in $AdS_5\times S_5$ in
\cite{Das:2004hy}.

To begin with note that the left-invariant  (super) current
defined as $J=g^{-1}dg$ satisfies the zero curvature equation
\begin{equation}
dJ+J\wedge J=0
\end{equation}
or explicitly
\begin{equation}
\partial_\mu J_\nu-\partial_\nu
J_\mu+[J_\mu,J_\nu]=0 \ .
\end{equation}
Using this equation we can express
the time component of the current $J_0$
as
\begin{equation}\label{D1j0}
\partial_1J_0+[J_1,J_0]\equiv
D_1J_0=\partial_0 J_1
\Rightarrow
J_0=D_1^{-1}(\partial_0 J_1) \ .
\end{equation}
where $D_1$ is defined by the first equality.

Although slightly unfamiliar, it turns out to be very convenient
to choose $J_1$ as a canonical variable and then to define the
conjugate momentum as the variation of the action with respect to
$\partial_0 J_1$ \cite{Korotkin:1997fi}. If we replace $J_0$ in
the action (\ref{Minaction}) with (\ref{D1j0})
 and then perform
the  variation with respect to $\partial_0 J_1$ we obtain
\begin{eqnarray}\label{PiJ}
 \Pi_J &&=\Pi^{(0)}+\Pi^{(1)}+
\Pi^{(2)}+\Pi^{(3)}=\nonumber \\
&&=-D_1^{-1} \left( D_1^{-1}(\partial_0 J_1)^{(2)}
+D_1^{-1}(\partial_0 J_1)^{(3)}+D_1^{-1}(\partial_0
J_1)^{(1)}\right.
\nonumber \\
&&\left. \quad -\frac{1}{2}J^{(3)}_1+ \frac{1}{2}J^{(1)}_1 -N_\mu
\cP^{\mu 0}-\hN_{\mu} \tP^{\mu 0}\right) \ ,
\end{eqnarray}
where we have used the fact that $$ \int d^2x \str [(D_1^{-1}G)F]=
-\int d^2x\str[G (D_1^{-1}F)] \ .
$$

We can then introduce the equal-time graded Poisson bracket that for
two classical observables $F,G$ depending on the phase super-space  variables
$Z^A\equiv J_1^A,\Pi_A$
 is defined as
\begin{eqnarray}
\pb{F,G} =(-1)^{|F||A|} \left[\frac{\partial^L F}{\partial Z^A}
\frac{\partial^L G}{\partial \Pi_A} -(-1)^{|A|}\frac{\partial^L
F}{\partial \Pi_A} \frac{\partial^L G}{\partial Z^A}\right] \ ,
\end{eqnarray}
where the superscript $L$ denotes left derivation. For the
components $J_1=J^A_1T_A$, $\Pi_J= \Pi^AT_A=K^{AB}\Pi_BT_A$, the above PB's read
\begin{equation}\label{canpb}
\pb{J^A_1(x),\Pi_B(y)}=(-1)^{|A|}
\delta^A_B\delta(x-y) \
\end{equation}
or explicitly
\begin{eqnarray}\label{pbe}
\pb{J^{\uc}_1(x),\Pi_{\ud}(y)} &=& \delta^{\uc}_{\ud}\delta(x-y) \ ,
\nonumber \\
\pb{J^{[\uc\ud]}_1(x),\Pi_{[\ue\uf]}(y)} &=&
\delta^{[\uc\ud]}_{[\ue\uf]} \delta(x-y) \ ,
\nonumber \\
\pb{J^{\alpha}_1(x),\Pi_{\beta}(y)} &=&
-\delta^\alpha_\beta\delta(x-y) \ ,
\nonumber \\
\pb{J^{\halpha}_1(x),\Pi_{\hbeta}(y)} &=&
-\delta^{\halpha}_{\hbeta}\delta(x-y) \ .
\end{eqnarray}

It is convenient to define  $\Pi^A$ as
\begin{equation}\label{PAB}
\Pi^A=K^{AB}\Pi_B \
\end{equation}
and to express $J_0^{A}$ as a function of the canonical variables
$J^A_1,\Pi^A$. With the help of (\ref{PiJ}) we get
\begin{eqnarray}\label{J0pij1}
&&J_0^\alpha= -(\partial_1\Pi^\alpha
+J^{[\uc\ud]}_1\Pi^{\beta}f_{[\uc\ud]\beta}^{\alpha}
+J^{\beta}_1\Pi^{[\uc\ud]} f_{\beta [\uc\ud]}^{\alpha}
+J^{\uc}_1\Pi^{\halpha}f_{\uc\halpha}^{\alpha}
+J^{\halpha}_1\Pi^{\uc}f_{\halpha\uc}^{\alpha})
-\frac{1}{2}J^\alpha_1 \ ,
\nonumber \\
&&J_0^{\halpha}= -(\partial_1\Pi^{\halpha}
+J^{[\uc\ud]}_1\Pi^{\hbeta}f_{[\uc\ud]\hbeta}^{\halpha}
+J^{\hbeta}_1\Pi^{[\uc\ud]} f_{\hbeta [\uc\ud]}^{\halpha}
+J^{\uc}_1\Pi^{\alpha}f_{\uc\alpha}^{\halpha}
+J^{\alpha}_1\Pi^{\uc}f_{\alpha\uc}^{\halpha})
+\frac{1}{2}J^{\halpha}_1 \ ,
\nonumber \\
&&J^{\uc}_0=-(\partial_1\Pi^{\uc}
+J^{[\uc\ud]}_1\Pi^{\uf}f_{[\uc\ud]\uf}^{\uc}+
J^{\uf}_1\Pi^{[\uc\ud]} f_{\uf [\uc\ud]}^{\uc}
+J^\alpha_1\Pi^\beta f_{\alpha\beta}^{\uc}+
J^{\halpha}_1\Pi^{\hbeta}f_{\halpha \hbeta}^{\uc}) \ ,
\nonumber \\
&&\Phi^{[\uc\ud]}=
\partial_1\Pi^{[\uc\ud]}+J^{[\ue\uf]}_1
\Pi^{[\ug\uh]} f_{[\ue\uf][\ug\uh]}^{[\uc\ud] }
+J^{\halpha}_1\Pi^\alpha f_{\halpha\alpha}^{[\uc\ud]}
+J^\alpha_1 \Pi^{\hbeta}f_{\alpha\hbeta}^{[\uc\ud]}
+J^{\ue}\Pi^{\uf}f_{\ue\uf}^{[\uc\ud]}-\nonumber \\
&& \qquad \quad -N^{[\uc\ud]}_\mu \cP^{\mu 0}- \hN^{[\uc\ud]}_\mu
\tP^{\mu 0} \ .
\end{eqnarray}

With (\ref{J0pij1}) in mind, we observe few important points.
Firstly, the expression $\Phi$ is the constraint that reflects
invariance of the action under local gauge $SO(4,1)\times SO(5)$
transformations. Secondly,  due to the fact that, contrary to the
standard GS action, the action (\ref{Minaction}) contains time
components of the currents
 $J^{\alpha}$, $J^{\halpha}$, it is not invariant under
 local $\kappa$ symmetry. As a result, in the present approach,
 the Hamiltonian analysis performed above
 does not generate the 'troublesome' fermionic
 constraints of the GS approach that
 cannot be covariantly  split into first  and second class,
the former being the generators of  $\kappa$ symmetry
 \cite{Das:2004hy}.  Yet, as we will momentarily see, the pure spinor
constraint could be viewed as the generator of local gauge
transformation of the $w$ and $\hat{w}$ conjugate ghosts.


\subsection{Graded Poisson brackets of the currents}
\label{GradedPB}

In this subsection we determine the graded algebra of Poisson
brackets of the currents using (\ref{pbe}) and also
(\ref{J0pij1}).

To begin with, note that by
 definition, the Poisson
bracket between currents with spatial
components is equal to zero
\begin{equation}
\pb{J^A_1(x),J^B_1(y)}=0 \ .
\end{equation}
Then it is rather straightforward to evaluate the Poisson brackets
of $J^A_0(x)$ and $J^B_1(y)$. Using (\ref{J0pij1}) and (\ref{pbe})
we get \be \pb{J^A_0(x),J^B_1(y)} = K^{AB} \partial_x \delta(x-y)
+ J^{C}_1(x) f_{C D}^{A} K^{DB} \delta(x-y) \ ,\ee or more
explicitly
\begin{eqnarray}
\pb{J^\alpha_0(x),J^{\beta}_1(y)}&=& J^{\uc}_1(x) f_{\uc
\halpha}^{\alpha} K^{\halpha\beta}\delta(x-y) \ ,
\nonumber \\
\pb{J^\alpha_0(x),J^{\hbeta}_1(y)}&=& K^{\alpha\hbeta}\partial_x
\delta(x-y)+J_1^{[\uc\ud]}(x)f_{[\uc\ud]\beta}^{\alpha}
K^{\beta\hbeta}\delta(x-y) \ ,
\nonumber \\
\pb{J^{\alpha}_0(x),J^{\uc}_1(y)}&=& J^{\halpha}_1(x)
f_{\halpha\ud}^\alpha K^{\ud\uc}\delta(x-y) \ ,
\nonumber \\
\pb{J^\alpha_0(x),J^{[\uc\ud]}_1(y)}&=& J^\beta_1(x)f_{\beta
[\ue\uf]}^\alpha K^{[\ue\uf][\uc\ud]}\delta(x-y) \ ,
\nonumber \\
\pb{J^{\halpha}_0(x),J^{\hbeta}_1(y)}&=& J^{\uc}_1(x) f_{\uc
\alpha}^{\halpha} K^{\alpha\hbeta}\delta(x-y) \ ,
\nonumber \\
\pb{J^{\halpha}_0(x),J^{\beta}_1(y)}&=& K^{\halpha\beta}\partial_x
\delta(x-y)+J_1^{[\uc\ud]}(x)f_{[\uc\ud]\hbeta}^{\halpha}
K^{\hbeta\beta}\delta(x-y) \ ,
\nonumber \\
\pb{J^{\halpha}_0(x),J^{\uc}_1(y)}&=& J^{\alpha}_1(x)
f_{\alpha\ud}^{\halpha} K^{\ud\uc}\delta(x-y) \ ,
\nonumber \\
\pb{J^{\uc}_0(x),J^\alpha_1(y)}&=&
J^{\halpha}_1(x)f_{\halpha\hbeta}^{\uc}
K^{\hbeta\alpha}\delta(x-y) \ ,
\nonumber \\
\pb{J^{\uc}_0(x),J^{\halpha}_1(y)}&=&
J^{\alpha}_1(x)f_{\alpha\beta}^{\uc} K^{\beta\halpha}\delta(x-y) \
,
\nonumber \\
\pb{J^{\uc}_0(x),J^{\ud}_1(y)}&=& K^{\uc\ud}\partial_x\delta(x-y)
+J_1^{[\uf\ug]}(x)f_{[\uf\ug]\ue}^{\uc}K^{\ue\ud}
\delta(x-y) \ , \nonumber \\
\pb{J^{\uc}_0(x),J^{[\uc\ud]}_1(y)}&=& J_1^{\ud}(x)f_{\ud
[\ue\uf]}^{\uc} K^{[\ue\uf][\uc\ud]}\delta(x-y) \ ,
\nonumber \\
\pb{J^{\halpha}_0(x),J^{[\uc\ud]}_1(y)}&=& J^{\hbeta}_1(x)f_{\hbeta
[\ue\uf]}^{\halpha} K^{[\ue\uf][\uc\ud]}\delta(x-y) \ .
\end{eqnarray}
The structure of these PB's deserves some comments.
 The first important feature
to notice is that they are not manifestly covariant w.r.t. two
dimensional worldsheet transformations. This is a consequence of
the non covariant  equal time
 Hamiltonian formalism.  Another feature is the presence of the
non ultra-local terms $\partial_x\delta(x-y)$.
 They arise as a result of our choice
of canonical variables. For a better understanding of this
approach,
 we will  perform in Appendix A a similar
analysis in the simplest case of a two dimensional free bosonic
theory. Finally and very importantly, the PB's respect the $Z_4$
grading dictated by the underlying $PSU(2,2|4)$ structure.

Along the same line, we can calculate the Poisson brackets between
the constraint $\Phi$ and the spatial currents $J^A_1$
\begin{eqnarray}
\pb{\Phi^{[\uc\ud]}(x),J^{[\ue\uf]}_1(y)} &=&
-\partial_x\delta(x-y)K^{[\uc\ud][\ue\uf]} -J_1^{[\ua\ub]}(x)
f_{[\ua\ub][\ug\uh]}^{[\uc\ud]} K^{[\ug\uh][\ue\uf]}\delta(x-y) \
,
\nonumber \\
\pb{\Phi^{[\ug\uh]}(x),J^{\uc}_1(y)}&=& -J^{\ud}_1(x) f_{\ud
\ue}^{[\ug\uh]}K^{\ue\uc}\delta(x-y) \ ,
\nonumber \\
\pb{\Phi^{[\uc\ud]}(x),J^\alpha_1(y)}&=& -J^\gamma_1(x)
f_{\gamma\hbeta}^{[\uc\ud]} K^{\hbeta\alpha} \delta(x-y) \ ,
\nonumber \\
\pb{\Phi^{[\uc\ud]}(x),J^{\halpha}_1(y)}&=&
-J^{\hgamma}_1(x)f_{\hgamma\beta}^{[\uc\ud]} K^{\beta\halpha}
\delta(x-y) \ ,
\end{eqnarray}
that explicitly show how   the left-invariant currents transform
under the (right) gauge transformations generated by
$\Phi^{[\uc\ud]}$.

More involved is the calculation of the  Poisson brackets
$\pb{J^A_0(x),J^B_0(y)}$. Again we have to resort on
(\ref{J0pij1}) and (\ref{pbe}) as well as on the (anti) symmetry
properties
\begin{eqnarray}\label{id}
f_{AB}^EK^{BF} = -(-1)^{|B||C|}f_{AC}^DK^{CE}K_{DB}K^{BF}=
-(-1)^{|E||F|}f_{AC}^FK^{CE}
\end{eqnarray}
and the graded Jacobi identities
\begin{eqnarray}\label{grJI}
0=(-1)^{|A||C|} f_{AD}^{ E}f_{BC}^{ D} +(-1)^{|B||A|}f_{BD}^{
E} f_{CA}^{  D}+(-1)^{|C||B|} f_{CD}^{ E}f_{AB}^{ D}=0 \  .
\end{eqnarray}
After straightforward though rather tedious calculations we obtain
\begin{eqnarray}
\pb{J^{\uc}_0(x),J^{\ud}_0(y)}&=& -(\Phi^{[\uf\ug]}+N^{[\uf\ug]}_\mu
\cP^{\mu 0}+ \hN^{[\uf\ug]}_\mu \tP^{\mu
0})(x)f_{[\uf\ug]\ue}^{\uc}K^{\ue\ud}
\delta(x-y)\ , \nonumber \\
\pb{J^{\uc}_0(x),J^{\alpha}_0(y)}&=&
(J^{\hbeta}_0-J_1^{\hbeta})(x)f_{\hbeta\hgamma}^{\uc}
K^{\hgamma\alpha} \delta(x-y) \ ,
\nonumber \\
\pb{J^{\uc}_0(x),J^{\halpha}_0(y)}&=&
(J^\beta_0+J_1^{\beta})(x)f_{\beta\gamma}^{\uc}K^{\gamma
\halpha}\delta(x-y) \ ,
\nonumber \\
\pb{J^{\uc}_0(x),\Phi^{[\ud\ue]}(y)}&=& -J^{\ud}_0(x)f_{\ud
[\uf\ug]}^{\uc}K^{[\uf\ug] [\ud\ue]}\delta(x-y)
 \
\end{eqnarray}
and
\begin{eqnarray}
\pb{J^\alpha_0(x),J^\beta_0(y)}&=&
(J^{\uc}_0-J^{\uc}_1)(x)f_{\uc\halpha}^\alpha
K^{\halpha\beta}\delta(x-y) \ ,
\nonumber \\
\pb{J^{\halpha}_0(x),J^{\hbeta}_0(y)}&=&
(J^{\uc}_0+J^{\uc}_1)(x)f_{\uc\alpha}^{\halpha}
K^{\alpha\hbeta}\delta(x-y)
 \ , \nonumber \\
\pb{J^{\alpha}_0(x),J^{\halpha}_0(y)}&=&
-(\Phi^{[\uc\ud]}+N^{[\uc\ud]}_\mu \cP^{\mu 0}+
\hN^{[\uc\ud]}_\mu\tP^{\mu 0})(x)f_{[\uc\ud]\gamma}^{\alpha}
K^{\gamma\halpha}\delta(x-y)
\ , \nonumber \\
\pb{J^{\alpha}_0(x),\Phi^{[\uc\ud]}(y)}&=& -J^{\gamma}_0(x)
f_{\gamma [\ue\uf]}^\alpha K^{[\ue\uf][\uc\ud]}\delta(x-y) \ ,
\nonumber \\
\pb{J^{\halpha}_0(x),\Phi^{[\uc\ud]}(y)}&=&
-J^{\hgamma}_0(x)f_{\hgamma [\ue\uf]}^{\halpha}
K^{[\ue\uf][\uc\ud]}\delta(x-y) \ .
\end{eqnarray}
Finally we also need the Poisson
brackets of the generators of the
gauge transformations
\begin{equation}\label{phiab}
\pb{\Phi^{[\uc\ud]}(x),\Phi^{[\ue\uf]}(y)}=
-\Phi^{[\ua\ub]}(x)f_{[\ua\ub][\ug\uh]}^{[\uc\ud]}
 K^{[\ug\uh][\ue\uf]}\delta(x-y) \ .
\end{equation}
Using the above form of the current algebra, we will momentarily
derive the classical Hamiltonian and the field equations, and
prove the nihilpotency and conservation of the classical BRST
charges. It is also clear that using the Poisson brackets given
above we can find the Poisson brackets between the chiral
components of the currents $J^A_\pm$ which are related to $J^A_z$
and ${J}^A_{\bar{z}}$ after Wick rotation. We will demonstrate a
simple instance of this calculation in the next section where we
will also calculate the Poisson brackets between the BRST charges
and some chiral currents. However due to the non-chirally split
structure of the algebra, for our purposes, it is more convenient
to work with the Poisson brackets given above.


\section{BRST charges and Hamiltonian}
\label{BRSTHamil}

In this section, we discuss the Hamiltonian and the BRST charges
together with their properties. We then derive the classical
canonical equations of motion in the next section.

As a first step, we need the action of the BRST charges on the
currents and ghost fields.  To begin with we express
$N^{[\uc\ud]}, \hN^{[\uc\ud]}$ using the ghosts and their
conjugate momenta. Since
\begin{eqnarray}\label{wpi}
w_{\mu\alpha}\cP^{\mu 0}&=&-\pi_{\alpha} \ , \qquad
w_{\mu\alpha}\cP^{\mu 1}=-\pi_{\alpha} \ ,
\nonumber \\
\hw_{\mu\halpha}\tP^{\mu 0}&=&-\hpi_{\halpha}  \ , \qquad
\hw_{\mu\halpha}\tP^{\mu 1}=\hpi_{\halpha}
\end{eqnarray}
we obtain
\begin{eqnarray}\label{Npilambda}
N_{\mu}^{[\uc\ud]} \cP^{\mu 0}&=&\pi_\beta K^{\beta\hbeta}
f_{\hbeta\alpha}^{[\uc\ud]}\lambda^\alpha\equiv N^{[\uc\ud]} \ ,
\nonumber \\
N_{\mu}^{[\uc\ud]} \cP^{\mu 1}&=&\pi_\beta K^{\beta\hbeta}
f_{\hbeta\alpha}^{[\uc\ud]}\lambda^\alpha=N^{[\uc\ud]} \ ,
\nonumber \\
\hN_{\mu}^{[\uc\ud]}\tP^{\mu 0}&=&
\hpi_{\hbeta}K^{\hbeta\beta}f_{\beta\hgamma}^{[\uc\ud]}
\hlambda^{\hgamma}\equiv \hN^{[\uc\ud]} \ ,
\nonumber \\
\hN_{\mu}^{[\uc\ud]}\tP^{\mu 1}&=&
-\hpi_{\hbeta}K^{\hbeta\beta}f_{\beta\hgamma}^{[\uc\ud]}
\hlambda^{\hgamma}=-\hN^{[\uc\ud]} \ ,
\end{eqnarray}
where $\pi,\lambda$ and $\hpi,\hlambda$ satisfy  the  canonical
Poisson brackets
\begin{equation}\label{canlpi}
\pb{\lambda^{\alpha}(x),\pi_\beta(y)}=
\delta^\alpha_\beta\delta(x-y) \ , \qquad
\pb{\hlambda^{\halpha}(x),\hpi_{\hbeta}(y)}=
\delta^{\halpha}_{\hbeta}\delta(x-y) \ .
\end{equation}
However one has to consider effect of the pure spinor constraints
on the system (\ref{purespinri}). Their presence implies that it
is
 natural to study the classical
dynamics of the ghosts system as the dynamics of a constrained
system. Using (\ref{wpi}) and (\ref{canlpi}), it is easy to see
that the constraints
\begin{equation}\label{Hc}
\Phi^{\uc}=\frac{1}{2}\lambda^\alpha \gamma_{\alpha\beta}^{\uc}\lambda^\beta \ ,
\quad \hPhi^{\uc}=\frac{1}{2}\hlambda^{\halpha}\gamma_{\halpha\hbeta}^{\uc}
\hlambda^{\hbeta} \ .
\end{equation}
generate the gauge transformations (\ref{locw}) since
\begin{equation}\label{Phiucpi}
\pb{\Phi^{\uc}(x), \pi_{\beta}(y)}=\gamma^{\uc}_{\beta\gamma}
\lambda^{\gamma}(y)
\delta(x-y)  \ ,
\pb{\hPhi^{\uc}(x), \hpi_{\hbeta}(y)}=\gamma^{\uc}_{\hbeta
\hgamma}\hlambda^{\hgamma}(y)
\delta(x-y)  \ .
\end{equation}
Using (\ref{Phiucpi}) we also
obtain
\begin{eqnarray}\label{PhiucpN}
\pb{\Phi^{\uc}(x),N^{[\ud\ue]}(y)}=
-\Phi^{\uh}(x)f_{\uh [\ua\ub]}^{\uc}
K^{[\ua\ub][\ud\ue]}\delta(x-y) \ ,
\nonumber \\
 \pb{\hPhi^{\uc}(x),\hN^{[\ud\ue]}(y)}=
-\hPhi^{\uh}(x)f_{\uh [\ua\ub]}^{\uc}
K^{[\ua\ub][\ud\ue]}\delta(x-y) \ .
\end{eqnarray}
In the same way we can show that
\begin{equation}\label{Phiucd}
\pb{\Phi^{\uc}(x),\Phi^{\ud}(y)}=0 \ , \quad
\pb{\hPhi^{\uc}(x),\hPhi^{\ud}(y)}=0 \ .
\end{equation}
This result implies that the pure spinor constraints are first
class.

Let us then consider the Poisson bracket of $\Phi^{[\uc\ud]}$ with
the ghost variables. Using the  explicit form of $\Phi^{[\uc\ud]}$
given in (\ref{J0pij1}) and also (\ref{Npilambda}) together with
(\ref{canlpi}) we obtain
\begin{eqnarray}\label{Philambda}
\pb{\Phi^{[\uc\ud]}(x),\lambda^\alpha(y)}&=& -
\lambda^\beta(x)
f_{\beta\hbeta}^{[\uc\ud]}K^{\hbeta\alpha}
\delta(x-y) \ ,
\nonumber \\
\pb{\Phi^{[\uc\ud]}(x),\hlambda^{\halpha}(y)}&=& -
\hlambda^{\hbeta}(x)
f_{\hbeta\beta}^{[\uc\ud]}K^{\beta\hbeta}\delta(x-y) \
\end{eqnarray}
that explicitly demonstrates that
 $\lambda$, $\hlambda$ transform
nontrivially under $SO(4,1)\times SO(5)$
gauge transformations.
Moreover, (\ref{Philambda}) also
implies
\begin{eqnarray}\label{Phicd}
\pb{\Phi^{[\uc\ud]}(x),
\Phi^{\ue}(y)}=\Phi^{\uf}(x)f_{\uf
[\ua\ub]}^{\ue}K^{[\ua\ub][\uc\ud]}
\delta(x-y) \ , \nonumber \\
\pb{\Phi^{[\uc\ud]}(x), \hPhi^{\ue}(y)}=\hPhi^{\uf}(x)f_{\uf
[\ua\ub]}^{\ue}K^{[\ua\ub][\uc\ud]} \delta(x-y) \ .
\end{eqnarray}
Then  (\ref{phiab}), (\ref{Phiucd})
and  (\ref{Phicd}) show that
$\Phi^{[\uc\ud]},\Phi^{\uc},\hPhi^{\uc}$
 consist of only first
class constraints. This fact will be important below.

For later purposes we here
determine the following Poisson brackets
\begin{eqnarray}
\pb{\Phi^{[\uc\ud]}(x),\pi_\beta(y)}
&=& K^{[\uc\ud][\ue\uf]}
f_{[\ue\uf]\beta}^\gamma \pi_\gamma(y)
\delta(x-y) \ , \nonumber \\
\pb{\Phi^{[\uc\ud]}(x),\hpi_{\hbeta}(y)}&=&
K^{[\uc\ud][\ue\uf]}f_{[\ue\uf]\hbeta}^{\hgamma} \hpi_{\hgamma}(y)
\delta(x-y)
\end{eqnarray}
and
\begin{eqnarray}
\pb{\Phi^{[\uc\ud]}(x),N^{[\ue\uf]}(y)}=
-N^{[\ug\uh]}(y)f_{[\ug\uh][\ua\ub]}^{[\uc\ud]}
 K^{[\ua\ub][\ue\uf]}\delta(x-y) \ ,
\end{eqnarray}
\begin{equation}
\pb{\Phi^{[\uc\ud]}(x),\hN^{[\ue\uf]}(y)}=
 -\hN^{[\ug\uh]}(y)
f_{[\ug\uh][\ua\ub]}^{[\uc\ud]}
K^{[\ua\ub][\ue\uf]}\delta(x-y) \ .
\end{equation}

\subsection{Classical BRST generators}
\label{classBRST}

We are ready to study  the action of the BRST charges on the
fundamental fields that appear in the action (\ref{Minaction}). As
shown in \cite{Berkovits:2000yr, Berkovits:2004xu,Kluson:2006wq}
the BRST charges take the form
\begin{eqnarray}
Q_R=\int dx \hlambda^{\halpha}
K_{\halpha\alpha}J^{\alpha}_\mu
\tP^{\mu 0}
=
-\int dx \hlambda^{\halpha}
K_{\halpha\beta}[J^\beta_0+J^\beta_1] \ ,
\nonumber \\
 Q_L=\int dx \lambda^{\alpha}K_{\alpha
 \hbeta}J^{\hbeta}_\mu \cP^{\mu 0}=
-\int dx \lambda^\alpha K_{\alpha\hbeta}
[J^{\hbeta}_0-J^{\hbeta}_1] \ .
\end{eqnarray}
Then using the Poisson brackets
determined in the previous section
we easily get
\begin{eqnarray}\label{QRL}
\pb{Q_R,J^{\uc}_1(y)}&=& -\hlambda^{\halpha}J^{\hbeta}_1(y)
f_{\halpha \hbeta}^{\uc} \ , \qquad \pb{Q_L,J^{\uc}_1(y)}=-
\lambda^{\alpha}J^{\beta}_1(y) f_{\alpha\beta}^{\uc} \ ,
 \nonumber \\
\pb{Q_R,J_0^{\uc}(y)}&=& -\hlambda^{\halpha} J^{\hbeta}_0(y)
f_{\halpha\hbeta}^{\uc} \ , \qquad
 \pb{Q_L,J^{\uc}_0(y)}=
-\lambda^\alpha
J^\beta_0(y) f_{\alpha\beta}^{\uc} \ ,
\nonumber \\
\pb{Q_R,J^\alpha_1(y)}&=& -\hlambda^{\halpha}J^{\uc}_1(y)
f_{\halpha\uc}^{\alpha} \ , \qquad \pb{Q_L,J^{\halpha}_1(y)}=
-\lambda^\beta J^{\uc}_1(y) f_{\beta\uc}^{\halpha} \ ,
\nonumber \\
\pb{Q_L,J^{\alpha}_1(y)}&=&
\partial_1\lambda^\alpha(y)+J_1^{[\uc\ud]}\lambda^\beta(y)
f_{[\uc\ud]\beta}^\alpha \equiv \nabla_1\lambda^\alpha(y)\ , \nonumber \\
\pb{Q_R,J^{\halpha}_1(y)}&=&
\partial_1\hlambda^{\halpha}(y)+
J_1^{[\uc\ud]}\hlambda^{\hbeta}(y)f_{[\uc\ud]\hbeta}^{\halpha}
\equiv \nabla_1 \hlambda^{\halpha}(y) \end{eqnarray} where
$\nabla_1 X^A = \partial_1 X^A +
J_1^{[\uc\ud]}X^B(y)f_{[\uc\ud]B}^{A}$, and also
\begin{eqnarray}\label{QRL1}
\pb{Q_R,J^\alpha_0(y)}&=& -\hlambda^{\halpha} J^{\uc}_0(y)
f_{\halpha\uc}^\alpha \ ,
\nonumber \\
\pb{Q_L,J^\alpha_0(y)} &=&
\lambda^\gamma(\Phi^{[\uc\ud]}+N^{[\uc\ud]}_\mu \cP^{\mu 0}+
\hN^{[\uc\ud]}_\mu \tP^{\mu 0})(y)f_{\gamma [\uc\ud]}^\alpha-
\nabla_1\lambda^\alpha(y) \ ,
\nonumber \\
\pb{Q_R,J^{\halpha}_0(y)}&=&
\hlambda^{\hgamma}(\Phi^{[\uc\ud]}+N^{[\uc\ud]}_\mu \cP^{\mu 0}
+\hN_\mu^{[\uc\ud]} \tP^{\mu 0})(y)f_{\hgamma [\uc\ud]}^\alpha+
\nabla_1 \hlambda^{\halpha}(y) \ ,
\nonumber \\
\pb{Q_L,J^{\halpha}_0(y)}&=& -\lambda^\alpha J^{\uc}_0(y)f_{\alpha
\uc}^{\halpha} \ .
\end{eqnarray}
It turns out that we will also need the following Poisson brackets
\begin{eqnarray}\label{QRL2}
\pb{Q_L,J^{[\uc\ud]}_1(y)}&=& \lambda^\alpha J_1^{\hgamma}(y)
f_{\alpha \hgamma }^{[\uc\ud]} \ ,
\nonumber \\
\pb{Q_R,J^{[\uc\ud]}_1(y)}&=& \hlambda^{\halpha}J_1^{\beta}(y)
f_{\halpha\beta}^{[\uc\ud]} \ .
\end{eqnarray}
The Poisson bracket between
BRST charges and ghost fields
can be easily worked out
using   (\ref{canlpi}) and
we obtain
\begin{eqnarray}\label{QRLlp}
\pb{Q_{(L,R)},\lambda^\alpha(y)}&=&
\pb{Q_{(L,R)},\hlambda^{\halpha}(y)}=0 \ ,
\nonumber \\
\pb{Q_L,\pi_{\alpha}(y)}&=&
-K_{\alpha\hbeta}[J_0^{\hbeta}-J_1^{\hbeta}](y) \ , \qquad
\pb{Q_R,\pi_{\alpha}(y)}=0 \ ,
\nonumber \\
\pb{Q_R,\hpi_{\halpha}(y)}&=&
-K_{\halpha\beta}[J^\beta_0+J^\beta_1](y) \ , \qquad
\pb{Q_L,\hpi_{\halpha}(y)}=0 \ .
\end{eqnarray}
In the same way
 we can  determine
the Poisson brackets between BRST charges and  $N^{[\uc\ud]} \ ,
\hN^{[\uc\ud]}$
\begin{eqnarray}\label{QRLlp1}
\pb{Q_L,N^{[\uc\ud]}(y)}&=&
[J_0^{\hbeta}-J_1^{\hbeta}]\lambda^\alpha(y)
f_{\hbeta\alpha}^{[\uc\ud]}  \ ,
\nonumber \\
\pb{Q_R,\hN^{[\uc\ud]}(y)}&=&
[J^\beta_0+J^\beta_1]\hlambda^{\halpha}(y)
 f_{\beta\halpha}^{[\uc\ud]}
 \  .
\end{eqnarray}

Before we conclude this section we would like to briefly discuss
the BRST transformations of the (light-cone) components of the
currents
\begin{equation}
J_\pm^A=\frac{1}{\sqrt{2}}
(J_0^A\pm J^A_1) \ .
\end{equation}
It is rather straightforward to calculate the action of the BRST
charges $Q_R,Q_L$ on the (chiral) currents $J^A_\pm$. For
illustration, let us consider the action of the charge $Q=Q_R+Q_L$
on the current $J_\pm^{\uc}$. Using (\ref{QRL}) we obtain
\begin{eqnarray}
\pb{Q,J^{\uc}_+(y)}=
-\hlambda^{\halpha}J^{\hbeta}_+(y) f_{\halpha
\hbeta}^{\uc}
-\lambda^{\alpha}J^{\beta}_+(y) f_{\alpha
\beta}^{\uc} \ ,
\nonumber \\
\pb{Q,J^{\uc}_-(y)}=-\lambda^\alpha J_-^\beta(y)
f_{\alpha\beta}^{\uc}-\hlambda^{\halpha}
J^{\hbeta}_-(y)f_{\halpha\hbeta}^{\uc} \ .
\end{eqnarray}
In the same way we can calculate the action of the BRST charge $Q$
on all remaining currents. Since the procedure is straightforward
we will not report it here. However we have to stress one
important point. It can be easily shown
 that the action of the BRST charges on the
chiral currents that in
the canonical formalism is defined
as the Poisson bracket between
BRST charge $Q$ and corresponding
current,
does not fully coincide with the
 BRST transformation of currents
given in \cite{Berkovits:2000yr}. This follows from the fact that
our calculation is based on Hamiltonian formalism that is not
manifestly covariant.
 Secondly, the
transformation of the currents given in \cite{Berkovits:2000yr} is
a combination of a BRST transformation and a gauge transformation.
Unfortunately it is not completely clear to us how these
transformations are related to the BRST transformations given
here.

\subsection{Hamiltonian}
\label{classHam}

At this point we are ready to determine the Hamiltonian for the
pure spinor string in $AdS_5\times S_5$. Using the supergroup
notation, we define the matter part of the Hamiltonian as
\begin{eqnarray}\label{Hmatt}
&& H_{matt} = \int dx \str (\partial_0 J_1\Pi-{\mL}_{matt}) =\int dx
\left( \frac{1}{2} \left[ J^{\uc}_0 J^{\ud}_0K_{\uc\ud}+
J^{\uc}_1J^{\ud}_1K_{\uc\ud}+J^{\alpha}_0
J^{\hbeta}_0 K_{\alpha\hbeta} \right. \right. \nonumber \\
&& \left. \left. + J^{\hbeta}_0J^{\alpha}_0K_{\hbeta\alpha}+
J^{\alpha}_1 J^{\hbeta}_1K_{\alpha\hbeta}+ J^{\hbeta}_1
J^{\alpha}_1K_{\hbeta\alpha} \right] +
N^{[\uc\ud]}K_{[\uc\ud][\ue\uf]}
J^{[\ue\uf]}_1-\hN^{[\uc\ud]}K_{[\uc\ud][\ue\uf]} J^{[\ue\uf]}_1
\right) \ . \nonumber \\
\end{eqnarray}
In the same way we define the ghost part of the Hamiltonian as
\begin{eqnarray}
 H_{ghost}= &&\int dx \left(\pi_\alpha
\partial_0\lambda^\alpha+
\hpi_{\halpha}\partial_0\hlambda^{\halpha}-\mL_{ghosts} \right)=
 \nonumber \\
&& \int dx\left(-\pi_{\alpha} \partial_1\lambda^\alpha
+\hpi_{\halpha}\partial_1\hlambda^{\halpha}+
N^{[\uc\ud]}K_{[\uc\ud][\ue\uf]}\hN^{[\ue\uf]}\right)
\end{eqnarray}
using the fact that
\begin{eqnarray}
N^{[\uc\ud]}_{\mu}\cP^{\mu\nu}K_{[\uc\ud][\ue\uf]}
\hN^{[\ue\uf]}_{\nu}=
-N^{[\uc\ud]}K_{[\uc\ud][\ue\uf]}\hN^{[\ue\uf]} \  .
\end{eqnarray}
Finally we  introduce the Hamiltonian that corresponds to the
$SO(4,1)\times SO(5)$ gauge symmetry constraints
and to the pure spinor constraints (\ref{Hc})
\begin{eqnarray}\label{Hcons}
H_{cons}=H_{coset}+H_{pure} \ ,
\nonumber \\
H_{coset}=\int dx \Gamma_{[\uc\ud]} \Phi^{[\uc\ud]} \ ,
\nonumber \\
H_{pure}=\int dx( \Gamma_{\uc}\Phi^{\uc}+
\hGamma_{\uc}\hPhi^{\uc}) \ ,
\end{eqnarray}
where $\Gamma_{[\uc\ud]}\ , \Gamma_{\uc} \ , \hGamma_{\uc}$  are
some {\it a priori} arbitrary functions of the phase space
variables $ (J_1^A,\Pi_A,\lambda,\hlambda, \pi,\hpi)$ \footnote{It
would be certainly interesting to perform a ``more symmetric"
analysis, whereby the  generalized BRST operators include the
constraints $\Phi^{[\uc\ud]},\Phi^{\uc}, \hPhi^{\uc}$, as
suggested in \cite{Chesterman:2002ey}. We leave this analysis to
future work.}. Then the total Hamiltonian is equal to
\begin{equation}\label{Htot}
H=H_{matt}+H_{ghost}+ H_{cons} \ .
\end{equation}
The general theory of  constrained systems requires that one
make sure that the time evolution of the constraints does not
generate any additional (secondary)
ones \cite{Henneaux:1992ig}.
Let us begin with $\Phi^{[\uc\ud]}$ and
prove that
\begin{equation}
\pb{\Phi^{[\uc\ud]}(x),H}\approx 0  \  ,
\end{equation}
where $\approx$ means that this Poisson bracket  vanishes on
constraint surface $\Phi^{[\uc\ud]}=0$.

Firstly, it can be explicitly
shown, using the Poisson
brackets given in the previous
section that
\begin{equation}
\pb{\Phi^{[\uc\ud]}(x),H_{matt}+ H_{ghost}}=0 \ .
\end{equation}
This result can be also
considered as  a  consequence of
the fact that $H_{matt}+H_{ghost}$ are
manifestly gauge invariant. On
the other hand the Poisson bracket of
$\Phi^{[\uc\ud]}$ with $H_{coset}$ is equal to
\begin{eqnarray}
\pb{\Phi^{[\uc\ud]}(x),H_{coset}} &=&
\int dy \pb{\Phi^{[\uc\ud]}(x),
\Gamma_{[\ue\uf]}(y)}\Phi^{[\ue\uf]}(y)
+\nonumber \\
&+& \Gamma_{[\ue\uf]} K^{[\ue\uf][\ug\uh]}
f_{[\ug\uh][\ua\ub]}^{[\uc\ud]}
\Phi^{[\ua\ub]}(x) \approx 0 \ ,
\end{eqnarray}
where we have used (\ref{phiab}). Finally, the Poisson  bracket
between $\Phi^{[\uc\ud]}$ and $H_{pure}$ can be easily calculated
with the help of (\ref{Phicd}) and we get
\begin{eqnarray}\label{Hpurecd}
\pb{\Phi^{[\uc\ud]}(x),H_{pure}} =
\int dy(\pb{\Phi^{[\uc\ud]}(x),\Gamma_{\ue}(y)}
\Phi^{\ue}(y)   +  \pb{\Phi^{[\uc\ud]}(x),
\hGamma_{\ue}(y)}\hPhi^{\ue}(y))+
\nonumber \\
\Gamma_{\ue}\Phi^{\uf}(x)f_{\uf
[\ua\ub]}^{\ue}K^{[\ua\ub][\uc\ud]}
+\hGamma_{\ue}\hPhi^{\uf}(x)f_{\uf
[\ua\ub]}^{\ue}K^{[\ua\ub][\uc\ud]} \approx 0 \ .
\end{eqnarray}
In other words the Poisson  bracket between $\Phi^{[\uc\ud]}$ and
$H$ vanishes on constraint surface and hence the time evolution of
$\Phi^{[\uc\ud]}$ does not generate additional secondary
constraint.

The situation is slightly more complicated in case of the pure
spinor constraints (\ref{Hc}). In fact, it is  easy to see, using
(\ref{PhiucpN}) that
\begin{equation}
\pb{\Phi^{\uc}(x),H_{matt}}\approx 0 \ , \quad
\pb{\hPhi^{\uc}(x),H_{matt}}\approx 0 \ .
\end{equation}
Moreover, we can also show
in the same way as in
(\ref{Hpurecd}) that the Poisson
bracket between pure spinor constraints
and $H_{coset}$ vanishes on constraint
surface. Finally, using (\ref{Phiucd})
we can show that
\begin{equation}
\pb{\Phi^{\uc}(x),H_{pure}}=
\pb{\hPhi^{\uc}(x),H_{pure}}=0 \ .
\end{equation}

On the other hand  the Poisson
brackets between $\Phi^{\uc},
\hPhi^{\uc}$ and $H_{ghost}$
are equal to
\begin{eqnarray}\label{phicgh}
\pb{\Phi^{\uc}(x),H_{ghost}}
=-\partial_1\lambda^\alpha
\gamma_{\alpha\beta}^{\uc}
\lambda^\beta(x)
-\Phi^{\ud}f_{\ud [\ue\uf]}^{\uc}
\hN^{[\ue\uf]}(x)=\nonumber \\
-\partial_1
\Phi^{\uc}(x)
-\Phi^{\ud}f_{\ud [\ue\uf]}^{\uc}
\hN^{[\ue\uf]}(x) \ ,
\nonumber \\
\pb{\hPhi^{\uc}(x),
H_{ghost}}=\partial_1\hlambda^{\halpha}
\gamma_{\halpha\hbeta}^{\uc}
\hlambda^{\hbeta}(x)
-\hPhi^{\ud}f_{\ud [\ue\uf]}^{\uc}
N^{[\ue\uf]}(x)=
\nonumber \\
\partial_1
\hPhi^{\uc}(x) -\hPhi^{\ud}f_{\ud [\ue\uf]}^{\uc} N^{[\ue\uf]}(x)
\ ,
\end{eqnarray}
where we have used (\ref{Phiucpi})
 and (\ref{PhiucpN}). We momentarily argue
that these expressions vanish
 along the constraints
 $\Phi^{\uc}=\hPhi^{\uc}=0$. It  is obvious
that this is true for the second terms on the second and the
fourth  line in (\ref{phicgh}). In order to clearly demonstrate
 that the first
term on the second line in (\ref{phicgh}) vanishes along the
constraints as well, note that it can be written as
\begin{equation}
\partial_1\Phi^{\uc}(x)=
\lim_{x'\rightarrow x}\frac{1}{(x'-x)}
(\Phi^{\uc}(x')-\Phi^{\uc}(x)) \ .
\end{equation}
In other words we can interpret this term as a difference of the
constraints at different points $x=x'$. Since the constraint
functions have to vanish for all $x$ it is now clear that this
difference vanishes as well. In the same way we can argue that the
first term on the fourth line in (\ref{phicgh}) vanishes on the
constraint surface $\hPhi^{\uc}=0$. In summary, the time evolution
of the pure spinor constraints does not generate new secondary
constraints.

\section{Equations of motions}
\label{Caneqs}

Using the form of the Hamiltonian (\ref{Htot}) and the known
Poisson brackets it is easy to determine the
 classical equations of motion
for currents and ghosts. We explicitly determine these equations
and show that they coincide with the equations of motion derived
in the Lagrangian formalism \cite{Berkovits:2000yr,
Kluson:2006wq,Vallilo:2003nx}, for an appropriate choice of the
gauge parameters,
\begin{equation}\label{eqm1}
\tP^{\mu\nu}
\nabla_\mu J^{(3)}_\nu
+
[J_\nu^{(3)},N_\mu]\cP^{\mu\nu}+
[J_\nu^{(3)},\hN_\mu]\tP^{\mu\nu}
=0 \ ,
\end{equation}
\begin{equation}\label{eqm2}
\cP^{\mu\nu}\nabla_\mu J^{(1)}_\nu
+
[J_\nu^{(1)},N_\mu]\cP^{\mu\nu}+
[J_\nu^{(1)},\hN_\mu]\tP^{\mu\nu}
=0 \ ,
\end{equation}
\begin{equation}\label{eqm3a}
\cP^{\mu\nu}
\nabla_\mu J^{(2)}_\nu
-\epsilon^{\mu\nu}
[J^{(1)}_\mu,J^{(1)}_\nu]
+[J_\nu^{(2)},N_\mu]\cP^{\mu\nu}+
[J_\nu^{(2)},\hN_\mu]\tP^{\mu\nu}
=0 \ ,
\end{equation}
\begin{equation}\label{eqm3b}
\tP^{\mu\nu}
\nabla_\mu J^{(2)}_\nu
+\epsilon^{\mu\nu}
[J^{(3)}_\mu,J^{(3)}_\nu]
+[J_\nu^{(2)},N_\mu]\cP^{\mu\nu}+
[J_\nu^{(2)},\hN_\mu]\tP^{\mu\nu}
=0 \ ,
\end{equation}
\begin{equation}\label{eqg1}
\cP^{\mu\nu}\nabla_\nu\lambda+
\cP^{\mu\nu}[\lambda,\hN_\nu]
=0 \ ,
\end{equation}
\begin{equation}\label{eqg2}
\tP^{\mu\nu}\nabla_\nu \hlambda
+\tP^{\mu\nu}[\hlambda,N_\nu]
=0 \ ,
\end{equation}
where
\begin{eqnarray}
\nabla_\nu J^{(i)}_\mu=
\partial_\nu J^{(i)}_\mu+
[J^{(0)}_\nu,J^{(i)}_\mu] \ ,
\nonumber \\
\nabla_\mu \lambda=
\partial_\mu \lambda+[J^{(0)}_\mu,\lambda] \ , \quad
\nabla_\mu \hlambda=
\partial_\mu \hlambda+
[J^{(0)}_\mu,\hlambda] \ ,
\end{eqnarray}
and where we also used the notations
defined in  (\ref{matnot}).

Let us now turn our attention onto the Hamiltonian formalism.
Recall that the time dependence of any classical observable is
governed by the equation
\begin{equation}
\partial_0 X=\pb{X,H} \ .
\end{equation}
We must also stress that we will write the resulting form of the
equations of motion that is valid along the constraints
$\Phi^{[\uc\ud]}=\Phi^{\uc}=\hPhi^{\ud}=0$.

Let us start with the
equation of motion for $\lambda^\alpha,
\hlambda^{\halpha}$.
Using
\begin{equation}
\pb{\lambda^\alpha(x),N^{[\uc\ud]}(y)}=
K^{\alpha\hbeta}f_{\hbeta\gamma}^{[\uc\ud]} \lambda^\gamma(x)
\delta(x-y) \
\end{equation}
and
\begin{equation}
\pb{\hlambda^{\halpha}(x),\hN^{[\uc\ud]}(y)}=
K^{\halpha\beta}f_{\beta\hgamma}^{[\uc\ud]} \hlambda^{\hgamma}(x)
\delta(x-y) \
\end{equation}
and also using (\ref{Htot}) we easily get
the equation of motion for
$\lambda^\alpha$
\begin{eqnarray}\label{eqgh}
\partial_0\lambda^\alpha=
\pb{\lambda^\alpha,H}= -\partial_1\lambda^\alpha
-J^{[\uc\ud]}_1\lambda^\gamma
 f_{ [\uc\ud]\gamma}^\alpha
-\hN^{[\uc\ud]} \lambda^\gamma
f_{[\uc\ud]\gamma }^\alpha
+\Gamma_{[\uc\ud]} \lambda^\gamma
f_{\gamma\hbeta}^{[\uc\ud]}
 K^{\hbeta\alpha} \ .
\end{eqnarray}
As we know $\Gamma_{[\uc\ud]}$ are arbitrary functions that
reflect the gauge invariance of the theory. However we can fix the
form of these parameters $\Gamma_{[\uc\ud]}$ in order to obtain
the form of the equation of motion that coincide with the
covariant equation (\ref{eqg1}).Using the fact that
\begin{equation}
\Gamma_{[\uc\ud]} \lambda^\gamma
f_{\gamma\hbeta}^{[\uc\ud]} K^{\hbeta\alpha}=
\Gamma^{[\uc\ud]}\lambda^\gamma f_{[\uc\ud]\gamma}^\alpha
\end{equation}
and by comparing (\ref{eqg1}) with
(\ref{eqgh}) we see that it
is natural to take
\begin{equation}\label{gpf}
\Gamma^{[\uc\ud]}=-J^{[\uc\ud]}_0 \ .
\end{equation}
In what follows we will assume the choice (\ref{gpf}) that in the
end will lead to the equivalence of the equations of motion
derived from the Hamiltonian formalism with the ones derived using
the Lagrangian formalism.

The equation of motion for $\hlambda$ can be easily derived as in
the case of the ghost $\lambda$ and it  coincides with
(\ref{eqg2}) with the help of (\ref{gpf}).

In the following, we will derive the equations of motion for
matter variables $J^{(i)}$ using the Poisson brackets derived in
the previous section and the matter Hamiltonian in (\ref{Hmatt}).

Let us start with the
equation of motion for
$J^{\uc}_1$
\begin{eqnarray}
\partial_0 J^{\uc}_1=
\pb{J^{\uc}_1,H}=
\nonumber \\
\partial_1 J^{\uc}_0+J^{[\ue\uf]}_1 J^{\ud}_0
f_{[\ue\uf]\ud}^{\uc}+J^{\halpha}_1 J^{\hbeta}_0
f_{\halpha\hbeta}^{\uc}+ J^{\alpha}_1 J^{\beta}_0
f_{\alpha\beta}^{\uc}-J^{[\ue\uf]}_0 J^{\ud}_1
f_{[\ue\uf]\ud}^{\uc}
\end{eqnarray}
that can be also written as
\begin{equation}\label{j21}
-\nabla_0 J_1^{(2)}+
\nabla_1 J_0^{(2)}
+[J^{(3)}_1, J^{(3)}_0]
+[J^{(1)}_1, J^{(1)}_0]
=0 \ .
\end{equation}
On the other hand the
equation of motion for
$J_0^{\uc}$ is more
involved and takes the form
\begin{eqnarray}
\partial_0 J^{\uc}_0=\pb{J^{\uc}_0,H}=
J_\nu^{\ud}N_\mu^{[\ue\uf]}
f_{\ud[\ue\uf]}^{\uc}
 \cP^{\mu \nu}
+J^{\ud}_\nu
\hN_\mu^{[\ue\uf]}f_{\ud[\ue\uf]}^{\uc}
 \tP^{\mu \nu}
+\nonumber \\
+\partial_1 J^{\uc}_1
+J_1^{[\ue\uf]} J_1^{\ud}f_{[\ue\uf]\ud}^{\uc}
 -J^{[\ue\uf]}_0J^{\ud}_0 f_{[\ue\uf]\ud}^{\uc} \ ,
\end{eqnarray}
or equivalently
\begin{equation}\label{j20}
\nabla_1 J^{(2)}_1-\nabla_0 J^{(2)}_0 +[J^{(2)}_\nu,
N_\mu]\cP^{\mu\nu}+ [J^{(2)}_\nu,\hN_\mu]\tP^{\mu\nu}
-[J_1^{(3)},J^{(3)}_0]+[J_1^{(1)}, J^{(1)}_0]=0 \ .
\end{equation}
 On the one hand,
if we  sum (\ref{j21}) with (\ref{j20}) we obtain the equation
that coincides with (\ref{eqm3a}). On the other hand, if we take
the difference of equations (\ref{j21}) and (\ref{j20}) we get an
equation that coincides with (\ref{eqm3b}).

Let us now consider the equation of motion for $J^{\alpha}_{1}$.
After some manipulations it can be written as
\begin{eqnarray}\label{j11}
-\nabla_0 J_0^{(1)}+ \nabla_1 J_1^{(1)}+ [J^{(3)}_0,J^{(2)}_1]
-[J^{(3)}_1,J^{(2)}_0]
+\nonumber \\
+[J^{(1)}_\nu,
 N_{\mu}]\cP^{\mu \nu}+
[J^{(1)}_\nu,\hN_{\mu}]\tP^{\mu \nu}=0 \ .
\end{eqnarray}
In the same way we can proceed with the equations of motion for
$J_0^{\alpha}$ that in the compact notation takes  the form
\begin{eqnarray}\label{j10}
\nabla_1 J_0^{(1)} -\nabla_0 J_1^{(1)} +[J^{(3)}_1, J^{(2)}_0]
-[J_0^{(3)},J_1^{(2)}]=0 \ .
\end{eqnarray}
It is easy to see that the if we add together  (\ref{j11}) with
(\ref{j10}) we derive
 the equation (\ref{eqm2}).
In the same way we can show that the equations of motion for
$J^{\halpha}_{1}, J^{\halpha}_0$ derived in the Hamiltonian
formalism
 imply the equation
(\ref{eqm1}).

\section{Conservation and nihilpotency
of the BRST charges}
\label{Consnihil}

In this section we will
show that the commutator
of the BRST charges
$Q_R,Q_L$ with the Hamiltonian
vanishes provided
 the dynamics is restricted to satisfy the local $SO(4,1)\times SO(5)$ constraint
$\Phi^{[\uc\ub]}=0$ and the pure spinor constraint for the ghost
fields. As a first step,
 we determine the Poisson bracket between
$H_{matt}$ and $Q_L$. Using the Poisson brackets given in
(\ref{QRL}), (\ref{QRL1}) and (\ref{QRL2})  we obtain
\begin{eqnarray}
\pb{Q_L,H_{matt}} &=& \int dx(-\partial_1\lambda^\alpha
K_{\alpha\hbeta} (J_0^{\hbeta}-J^{\hbeta}_1)+
(J_0^{\hbeta}-J^{\hbeta}_1) \lambda^\gamma
f_{\hbeta\gamma}^{[\uc\ud]}
K_{[\uc\ud][\ue\uf]}\hN^{[\ue\uf]} \nonumber \\
&+& \lambda^\gamma \Phi^{[\uc\ud]} f_{\gamma [\uc\ud]}^\alpha
K_{\alpha\hbeta}J^{\hbeta}_0+ \lambda^\gamma N^{[\uc\ud]}
f_{[\uc\ud]\gamma}^\alpha
K_{\alpha\hbeta}(J_0^{\hbeta}+J^{\hbeta}_1)) \ .
\end{eqnarray}

On the other hand
 the Poisson bracket of
$Q_L$ with $H_{ghost}$
can be easily worked out
using (\ref{QRLlp}) and (\ref{QRLlp1})
and  we get
\begin{eqnarray}
\pb{Q_L,H_{ghost}} &=& \int dx (\partial_1\lambda^\alpha
K_{\alpha\hbeta}[J_0^{\hbeta}-
J_1^{\hbeta}]-[J^{\hbeta}_0-J_1^{\hbeta}] \lambda^\alpha
f_{\hbeta\alpha}^{[\uc\ud]} K_{[\uc\ud][\ue\uf]}\hN^{[\ue\uf]}) \
.
\end{eqnarray}

Finally the Poisson bracket of $Q_L$
with $H_{coset}$ is equal
to
\begin{equation}
\pb{Q_L,H_{coset}}= \int dx\pb{Q_L,\Gamma_{[\uc\ud]}(x)}
\Phi^{[\uc\ud]}(x) \ ,
\end{equation}
where we have used the fact that
$\pb{Q_L,\Phi^{[\uc\ud]}(x)}=0$.
In the same way we can show that
\begin{equation}
\pb{Q_L,H_{pure}}\approx 0  \ .
\end{equation}
Collecting all these results  we obtain that the commutator of $Q_L$
with $H$ is equal to
\begin{eqnarray}\label{QlH}
\pb{Q_L,H}&=&\int dx(\lambda^\gamma f_{\gamma [\uc\ud]}^\alpha
K_{\alpha\hbeta}J^{\hbeta}_0+
\pb{Q_L,\Gamma_{[\uc\ud]}(x)})\Phi^{[\uc\ud]}(x)
\nonumber \\
&+& \int dx\lambda^\gamma N^{[\uc\ud]} f_{[\uc\ud]\gamma}^\alpha
K_{\alpha\hbeta}(J_0^{\hbeta}+J^{\hbeta}_1)\ .
\end{eqnarray}
The expression on the first line in (\ref{QlH}) is proportional to
$\Phi^{[\uc\ud]}$ that is
 zero on the constraint surface.
On the other hand using the explicit form of $N^{[\uc\ud]}$ we can
rewrite the expression on the second line  in (\ref{QlH}), omitting
the factor $K_{\alpha\hbeta}(J_0-J_1)^{\hbeta}$, as
\begin{eqnarray}
\lambda^\gamma N^{[\uc\ud]} f_{[\uc\ud]\gamma}^\alpha = \pi_\beta
K^{\beta\hbeta}f_{\hbeta \delta}^{[\uc\ud]} \lambda^\delta
f_{[\uc\ud]\gamma}^\alpha \lambda^\gamma= \frac{1}{2}\pi_\beta
K^{\beta\hbeta} f_{\hbeta \uc}^\alpha (\lambda^\delta
f_{\delta\gamma}^{\uc} \lambda^{\gamma}) \ ,
\end{eqnarray}
where in the final step we have used the  generalized Jacobi
identity (\ref{grJI}). However since $f_{\delta\gamma}^{\uc} = 2
(\gamma^{\uc})_{\delta\gamma}$ we obtain that the BRST charge
$Q_L$ is conserved on the constraint surface
\begin{equation}
\Phi^{[\uc\ud]}=\Phi^{\uc}=0 \ .
\end{equation}

In the same way we can calculate
the Poisson bracket of $Q_R$ with $H$ and
we obtain

\begin{eqnarray}\label{QRH}
\pb{Q_R,H}&=&\int dx (-J^{\alpha}_0\hlambda^{\hgamma}(x)
K_{\alpha\halpha} f_{\hgamma [\uc\ud]}^{\halpha}
+\pb{Q_R,\Gamma_{[\uc\ud]}(x)})\Phi^{[\uc\ud]}(x)
\nonumber \\
&-&\int dx \hN^{[\uc\ud]} f_{[\uc\ud]\hgamma }^{\halpha}
\hlambda^{\hgamma} K_{\halpha \alpha}(J^\alpha_0+J^\alpha_1) \ .
\end{eqnarray}
The expression on the first line in (\ref{QRH}) is again
proportional to the constraint $\Phi^{[\uc\ud]}$ and hence it
vanishes on the constraint surface
$\Phi^{[\uc\ud]}=0$.
 On the other hand the
expression on the second line is
proportional to
\begin{eqnarray}
-\hN^{[\uc\ud]} f_{[\uc\ud]\hgamma }^{\halpha} \hlambda^{\hgamma}
K_{\halpha \alpha}&=& -\pi_{\gamma}K^{\gamma\hdelta} f_{\hdelta
\hbeta}^{[\uc\ud]} \hlambda^{\hbeta}
f_{[\uc\ud]\hgamma}^{\halpha}\hlambda^{\hgamma}=
\nonumber \\
&=& \frac{1}{2}\pi_{\gamma}K^{\gamma\hdelta}
f_{\hdelta\uc}^{\halpha}( \hlambda^{\hgamma}
f_{\hgamma \hbeta}^{\uc}
\hlambda^{\hbeta})\sim \hPhi^{\uc}
\end{eqnarray}
and we see that the Hamiltonian `commutes' or, rather, is in
involution with $Q_R$ along the constraints. In other words we
have shown that the BRST charges are conserved as expected for any
generator of a global symmetries.

It is also important to
prove that the BRST charges are
nihilpotent at least on the constraint
surfaces
$\Phi^{[\uc\ud]}=\Phi^{\uc}=
\hPhi^{\uc}=0$.
In other words we have to show
 that the Poisson
brackets between $Q_R,Q_L$  vanish
or they are proportional to
generators of gauge transformations.
In fact, using the known form of
the Poisson bracket between BRST
generator $Q_L$ and the currents
$J^{A}$ we easily obtain
\begin{eqnarray}
\pb{Q_L,Q_L}&=& -\int dx \lambda^{\alpha}K_{\alpha\hbeta}
\left(\pb{Q_L,J^{\hbeta}_0(x)} -\pb{Q_L,J^{\hbeta}_1(x)}\right)=
\nonumber \\
&=& \int dx \lambda^\alpha K_{\alpha\hbeta}f^{\hbeta}_{\gamma\uc}
\lambda^\gamma [J^{\uc}_0-J^{\uc}_1] \ .
\end{eqnarray}
Using the fact that
\begin{equation}
\lambda^{\alpha}K_{\alpha\hbeta}
f^{\hbeta}_{\gamma\uc}\lambda^\gamma=
\lambda^\gamma
f_{\gamma\alpha}^{\ud} \lambda^\alpha K_{\ud\uc}
\sim \Phi^{\ud}
\end{equation}
we obtain the result
 that  the BRST charge $Q_L$ is
nihilpotent as a
consequence of the pure
spinor constraint (\ref{Hc}).
 We would like to stress that our proof
that $Q_L$ is nihilpotent is valid
even if all fields are off-shell.
It only relies on the local
$SO(4,1)\times SO(5)$ and pure spinor
constraints.

In case of $Q_R$ we proceed in the same way and we find that $Q_R$
is nihilpotent as well.

Finally, we can calculate the Poisson bracket between $Q_R$ and
$Q_L$
\begin{eqnarray}\label{qlqr}
\pb{Q_L,Q_R}&=& -\int dx \hlambda^{\halpha}
K_{\halpha\alpha}[\pb{Q_L,J^{\alpha}_0(x)}+
\pb{Q_L,J^{\alpha}_1(x)}]=
\nonumber \\
&=&-\int dx \hlambda^{\halpha}K_{\halpha\alpha}
\lambda^\gamma(\Phi^{[\uc\ud]}+
N^{[\uc\ud]}+ \hN^{[\uc\ud]})
f_{\gamma [\uc\ud]}^\alpha \  .
\end{eqnarray}
It is convenient to rewrite the term proportional to
$N^{[\uc\ud]}$ as
\begin{eqnarray}
-\lambda^\gamma N^{[\uc\ud]} f_{\gamma [\uc\ud]}^\alpha=
-\frac{1}{2}\lambda^\gamma f_{\gamma\delta}^{\uc} \lambda^\delta
f_{\uc\hbeta}^{\alpha}K^{\hbeta\beta}\pi_\beta \sim \Phi^{\uc}  \
\end{eqnarray}
and we   see that it vanishes
on the constraint surface $\Phi^{\uc}=0$.
In the same way we can show
\begin{eqnarray}
-\hlambda^{\halpha} K_{\halpha\alpha}\hN^{[\uc\ud]} f_{\gamma
[\uc\ud]}^\alpha= -\frac{1}{2}\hlambda^{\halpha}
f_{\halpha\hdelta}^{\uc}\hlambda^{\hdelta}
f_{\beta\uc}^{\hbeta}\hpi_{\hgamma}
K^{\hgamma\beta}K_{\hbeta\alpha}\sim \hPhi^{\uc}
\end{eqnarray}
that vanishes on the constraint
surface $\hPhi^{\uc}=0$.
Finally, the first term in
(\ref{qlqr}) is proportional to
$\Phi^{[\uc\ud]}$ and hence
it vanishes on the constraint
surface $\Phi^{[\uc\ud]}=0$.

Let us summarize the
results presented in this section. We have
shown that the Poisson brackets
between BRST generators vanish on
the constraint surface.
It is important that this result holds
without assuming that the
fundamental fields obey the equations of
motion. We also  hope that
this  result can be considered as an
additional support to the  analysis performed in
\cite{Berkovits:2004xu}.
It would be certainly very interesting to
extend this analysis to
the full quantum theory and further explore
the consequence of the
non-chiral splitting of the currents.

\section{Global currents and integrability}
\label{Globcurrint}

We would now like to study the classically conserved local
currents, that generate global $PSU(2,2|4)$ transformations, and
their non-local extensions, whose conservation strongly supports
classical integrability of the theory
\cite{Bena:2003wd,Dolan:2004ps,Vallilo:2003nx,Alday:2003zb,
Das:2004hy,Alday:2005gi}, within the present approach.

In the covariant pure spinor formalism the problem has been
studied by Vallilo \cite{Vallilo:2003nx}. One starts with a new
set of left-invariant currents $\hat{J}(u)$ satisfying the
flatness condition \be d\hat{J} + \hat{J} \wedge \hat{J} = 0 \ee
for any value of the spectral parameter $u$ with the 'initial'
condition $\hat{J}(0)= J = g^{-1} dg$. Making an ansatz of the
form 
\bea \hat{J}_\mu (u) = J_\mu  &+& \frac{1}{2}{\cP}_{\mu\nu} [
a(u)J^{\nu(2)} + b(u) J^{\nu (1)} + c(u) J^{\nu
(3)} + \tilde{d}(u) \hN^\nu ] \nn \\
&+& \frac{1}{2}\tilde{\cal P}_{\mu\nu} [ \tilde{a}(u)J^{\nu (2)} +
\tilde{b}(u) J^{\nu (1)} + \tilde{c}(u) J^{\nu(3)} + d(u)
N^\nu ] \eea and imposing flatness, using flatness of $J_\mu(0)$
and the classical field equations, derived above in a Hamiltonian
form or in \cite{Vallilo:2003nx,Berkovits:2004xu, Kluson:2006wq}
in a Lagrangian form, one gets\footnote{Our spectral parameter $u$
is related to the spectral parameter $\mu$ of
\cite{Vallilo:2003nx} by $\mu= e^u$.
Note also that we have chosen one particular solution from
the ones found in \cite{Vallilo:2003nx} in order to 
obey the initial condition $\hat{J}_\mu(0)=J_\mu$. 
It is remarkable that the classical theory admits the same two
one-parameter families of
flat currents if one sets the contribution of the pure spinor
ghost $N$ to zero.} \bea
 && a = e^u - 1 \qquad \tilde{a} = e^{-u} - 1 \nn
\\
&& b =  e^{3u/2} - 1 \qquad \tilde{b} =  e^{-u/2} - 1 \nn
\\
&& c =  e^{u/2} - 1 \qquad \tilde{c} =  e^{-3u/2} - 1 \nn
\\
&& d =  e^{2u}-1 \qquad \tilde{d} = 
e^{-2u}-1 \eea

so that eventually \begin{eqnarray}
 \hat{J}_\mu(u) =J_\mu+ ( \eta_{\mu\nu} (\cosh{u} -1)+
\epsilon_{\mu\nu} \sinh{u}) J^{\nu(2)}+
\nonumber \\
( \eta_{\mu\nu} (\cosh{u}e^{u/2} -1)+
\epsilon_{\mu\nu} \sinh{u}e^{u/2}) J^{\nu(1)}
+\nonumber \\
( \eta_{\mu\nu} (\cosh{u} e^{-u/2}-1)+
\epsilon_{\mu\nu} \sinh{u} e^{-u/2}) J^{\nu(3)}
 +\nonumber \\
+\sinh u e^{u}\tP_{\mu\nu} N^{\nu}
-\sinh ue^{-u}\cP_{\mu\nu}\hN^{\nu} \ . 
\nonumber \\ 
\end{eqnarray}

Flatness of the current $\hat{J}$ implies integrability for any $u$
of the equation \be \hat{D}_\mu \chi = 0 \ee where $\hat{D}_\mu
=\pa_\mu + \hat{J}_\mu$. It turns out to be convenient to exploit
the combination \be \epsilon_{\mu\nu} \pa^\nu \chi = -
\epsilon_{\mu\nu} \hat{J}^\nu \chi + u \pa_\mu \chi + u \hat{J}_\mu
\chi \quad . \ee

Setting $\hat{A}_\mu(u)= \hat{J}_\mu (u) - J_\mu = u g^{-1} a_\mu(u)
g$ (since $\hat{A}_\mu(0)=0$) one gets \be \epsilon_{\mu\nu} \pa^\nu
(g \chi) = u  \pa_\mu (g \chi) + (u^2 a_\mu(u) - u \epsilon_{\mu\nu}
a^\nu(u)) (g\chi) \quad .\ee

Expanding $\chi$ and $a_\mu$ in powers of $u$ around $u=0$, one
gets \be \epsilon_{\mu\nu} \pa^\nu (g \chi_{n}) =
\pa_\mu(g\chi_{n-1}) - \sum_{k=0}^{n-1} [\epsilon_{\mu\nu}
a^\nu_{k} - a_{\mu,k-1}] (g \chi_{n-k-1}) \ee The lowest order
yields \be n=0 \qquad \pa^\nu (g \chi_{0}) = 0 \ee that implies
$\chi_{0}= C g^{-1}$, where $C$ is a constant that we can set to
$C=1$ henceforth for simplicity. Plugging the latter in the second
equation yields
 \be n=1 \quad \epsilon_{\mu\nu} \pa^\nu (g \chi_1)
= - \epsilon_{\mu\nu} a^\nu_{0} 
\ee 
which in turn implies that
$j_{\mu, 0} = \epsilon_{\mu\nu} a^\nu_{0} $ is a classically
conserved local current. In the pure spinor approach one finds
 \be
j_{\mu, 0} = g \left[J_\mu^{(2)} + J^{(1)}_\mu + J^{(3)}_\mu +
{1\over 2} \epsilon_{\mu\nu} (J^{\nu(1)}- J^{\nu(3)}) + 
\tP_{\mu\nu} N^{\nu}+\cP_{\mu\nu}\hN^\nu)
\right]g^{-1} \ee Notice the difference w.r.t. the GS approach
where \be j^{GS}_{\mu, 0} = g \left[J_\mu^{(2)} + {1\over 2}
\epsilon_{\mu\nu} (J^{\nu(1)}_\mu - J^{\nu(3)}) \right]g^{-1} \ee
in addition to the pure spinor contribution, absent in the GS
approach, there is also an extra contribution in $J^{(1)}_\mu$ and
$J^{(3)}_\mu$ since they appear in the kinetic term and not only
in the WZ term, as required by $\kappa$ symmetry which is instead
fixed in the pure spinor approach.

The components  $j^A_{\mu, 0}= Str(T^A j_{\mu, 0})$ of the conserved
currents are expected to satisfy classical graded Poisson brackets
encoding the structure of the global $PSU(2,2|4)$ algebra.

As anticipated the procedure can be pushed forward to identify the
non-local currents.
 The first one arises at the next order where one
 finds
 \be j_{1\mu} = \epsilon_{\mu\nu} \pa^\nu (g \chi_{2}) =  -
\epsilon_{\mu\nu} (a^\nu_{0} g \chi_1 + a^\nu_{1}) \ee where \be
a_{\mu,1} = g [J_\mu^{(2)} + {5\over 8}(J^{(1)}_\mu + J^{(3)}_\mu) +
{1\over 2} \epsilon_{\mu\nu}(J^{\nu(1)}- J^{\nu(3)}) + \tP_{\mu\nu}N^{\nu}
+\cP_{\mu\nu}\hN^\nu]
g^{-1}\ee and \be g \chi_1 = - {1\over \pa^2} (\pa_\mu a_0^\mu) =
{1\over \pa^2} (\epsilon^{\mu\nu} \pa_\mu j_{0,\nu}) \ee so that \be
j_{1\mu} = j_{1\mu} {1\over \pa^2} (\epsilon^{\lambda\nu}
\pa_\lambda j_{0,\nu}) - \epsilon_{\mu\nu} a^\nu_{1} \ee and so on.

The classically conserved non local currents generate a Yangian
that has been studied for instance in \cite{Dolan:2004ps,
Agarwal:2004sz}.

\section{Conclusions}
\label{Conclus}

The present investigation has been  devoted to a classical
Hamiltonian analysis of the type IIB superstring on $AdS_5 \times
S^5$ in the pure spinor approach. Following \cite{Faddeev:1987ph,
Korotkin:1997fi}, we have taken the spatial components of the
(super)currents as canonical variables. In particular, we have
computed the classical graded Poisson brackets of the
left-invariant (super)currents and identified the first class
constraints associated to the gauging of $SO(4,1)\times SO(5)$. We
have then studied the properties of the BRST generators and the
Hamiltonian that governs the  dynamics of the system compatibly
with the local $SO(4,1)\times SO(5)$ and pure spinor constraints.
Contrary to the standard GS approach, whereby fermionic
constraints are both first and second class, the former being
associated to local $\kappa$ symmetry, the latter to the Dirac
constraint, all the constraints we have found are first class and
can be interpreted as generators of local symmetries. They appear
in the classical Hamiltonian via suitable Lagrange multipliers.
For a natural choice of the latter, we have satisfactorily shown
equivalence of the canonical equations of motion with the
covariant ones. Finally we have briefly discussed the global
symmetries and the issue of integrability within the present
framework.

It would be very interesting to further study  the structure of
the classical global algebra, that includes the global
$PSU(2,2|4)$ symmetry, and its representations. A crucial step
towards understanding the structure and classifying the classical
string configurations (``motions'') is determining the action of
the currents on the fundamental fields, either the coset
representative $g$ or the spatial components of the left-invariant
(super)currents, and ghosts. The latter are inescapably tangled
with the `matter' fields due to their non trivial transformations
under space-time symmetries. One could then tackle the much harder
issue of quantizing the string in this background and, in
particular, finding the spectrum of excitations beyond the
``massless'' supergravity states.

\section*{Acknowledgements}
We would like to thank A.~Das, P.~A.~Grassi, H.~Samtleben,
Ya.~Stanev for useful discussions. During completion of this work
M.B. was visiting the Galileo Galilei Institute for Theoretical
Physics of Arcetri (FI), INFN is acknowledged for hospitality and
support. M.B. would also like to thank the organizers
(C.~Angelatonj, E.~Dudas, T.~Gherghetta and A.~Pomarol) and the
participants to the workshop "Beyond the Standard Model" for
creating a stimulating environment.  This work was supported in
part by INFN, by the MIUR-COFIN contract 2003-023852, by the EU
contracts MRTN-CT-2004-503369 and MRTN-CT-2004-512194, by the
INTAS contract 03-516346 and by the NATO grant PST.CLG.978785. The
work of J.K. is also  supported in part by the Czech Ministry of
Education under Contract No. MSM 0021622409.

\newpage

\section*{Appendix A:
Properties of $PSU(2,2|4)$}\label{AppApsu}

In this Appendix, we briefly review the properties of the
superalgebra $psu(2,2|4)$, for more details we recommend the
papers \cite{Berkovits:2000yr,Berkovits:2004xu,
Metsaev:1998it,Das:2004hy}.

The generators of $psu(2,2|4)$ satisfy the graded commutation
relations

\begin{equation}
T_AT_B-(-1)^{|A||B|}T_BT_A= f_{AB}^{C}T_C \ ,
\end{equation}

The (super)index $A=(c,[cd],c',[c'd'],\alpha,\halpha)$ runs over
the tangent space indices of the super-Lie algebra of
$PSU(2,2|4)$, so that $(c,[cd])$ with $c, d =0,\dots,4$  describe
the $SO(4,2)$ isometries of $AdS_5$ and $(c',[c'd'])$ with $c',
d'=5\dots,9$ describe the $SO(6)$ isometries of $S^5$. We also
preserve the notation  $\alpha$ and $\halpha$ for the two
16-component Majorana-Weyl spinors. Finally, $\uc$ stands either
for $c$ or $c'$ (10 (pseudo)translations). In the same way
$[\uc\ud]$ stands either for $[cd]$ or for $[c'd']$ (10+10
(pseudo)rotations').

The non-vanishing structure constants $f_{AB}^C$ are
\begin{eqnarray}
&&f^{\uc}_{\alpha\beta}=2\gamma^{\uc}_{\alpha\beta}\ , \quad
f^{\uc}_{\halpha\hbeta}=2\gamma^{\uc}_{\halpha
\hbeta} \ , \nonumber \\
&&f^{[\ue\uf]}_{\alpha\hbeta}=f^{[\ue\uf]}_{\hbeta
\alpha}=(\gamma^{ef})_{\alpha}^{\gamma} \delta_{\gamma\hbeta} \ ,
\quad \ f^{[e'f']}_{\alpha\hbeta}=f^{[e'f']}_{\hbeta\alpha}=
-(\gamma^{e'f'})_{\alpha}^\gamma \delta_{\gamma\hbeta} \ ,
\nonumber \\
&&f^{\hbeta}_{\alpha\uc}=-f_{\uc\alpha}^{\hbeta}=
\frac{1}{2}(\gamma_{\uc})_{\alpha\beta}\delta^{\beta\hbeta} \ ,
\quad f_{\halpha\uc}^{\beta}=-f^{\beta}_{\uc\halpha}=
-\frac{1}{2}(\gamma_{\uc})_{\halpha\hbeta}\delta^{\beta\hbeta} \ ,
\nonumber \\
&&f^{[ef]}_{cd}=\frac{1}{2} \delta_c^{[e}\delta_d^{f]} \ , \quad
f_{c'd'}^{[e'd']}=-\frac{1}{2}\delta_{c'}^{[e'} \delta_{d'}^{f']}
\ , \quad f_{[\uc\ud]\ue}^{\uf}=-f_{\ue[\uc\ud]}^{\uf}=
\eta_{\ue[\uc}\delta_{\ud]}^{\uf} \ , \nonumber \\
&&f_{[\uc\ud][\ue\uf]}^{[\ug\uh]}=\frac{1}{2}
\left(\eta_{\uc\ue}\delta_{\ud}^{[\ug}\delta_{\uf}^{\uh]}
-\eta_{\uc\uf}\delta_{\ud}^{[\ug}\delta_{\uc}^{\uh]}
+\eta_{\ud\uf}\delta_{\uc}^{[\ug}\delta_{\ue}^{\uh]}
-\eta_{\ud\ue}\delta_{\uc}^{[\ug}\delta_{\uf}^{\uh]}\right) \ ,
\nonumber \\
&& f_{[\uc\ud]\alpha}^{\beta}=-f_{\alpha[\uc\ud]}^{\beta}=
\frac{1}{2}(\gamma_{\uc\ud})_\alpha^\beta \ , \quad
f_{[\uc\ud]\halpha}^{\hbeta}=-f_{\halpha[\uc\ud]}^{\hbeta}=
\frac{1}{2}(\gamma_{\uc\ud})_{\halpha}^{\hbeta} \ .
\end{eqnarray}

The graded-symmetric Cartan-Killing supermetric
\begin{equation}\label{Met} K_{AB}=\str
(T_AT_B)=(-1)^{|A||B|}K_{BA} \  ,
\end{equation}
with $|A|=1$ if $A$ is associated to a Grassmann odd generator and
$|A|=0$ if it is Grassmann even.

An essential feature of the superalgebra $psu(2,2|4)$ is that it
admits  $\mathbf{Z}_4$ automorphism $\Omega$ such that the
condition $\Omega(\mH)=\mH$ determines the maximal subgroup
$SO(4,1)\times SO(5)$ that has to be quotiented in the definition
of the coset.

The $\mathbf{Z}_4$ authomorphism $\Omega$ allows us to  decompose
the superalgebra ${\cal G}$ as
\begin{equation}
{\cal G} =\mH_0\oplus\mH_1\oplus\mH_2\oplus \mH_3 \ ,
\end{equation}
where $\mH_p$ denotes the eigenspace of $\Omega$  such that if
$\mathbf{h}_p\in \mH_p$ then
\begin{equation}\label{Omegaav}
\Omega(\mathbf{h}_p)=i^p\mathbf{h}_p \ .
\end{equation}
As we have argued above $\Omega(\mathbf{h}_0)= \mathbf{h}_0$
determines $\mH_0= SO(4,1)\times SO(5)$. $\mH_2$ includes the
remaining bosonic generators of the superalgebra, while
$\mH_1,\mH_3$ consist of the fermionic generators of the algebra.
The authomorphism $\Omega$ also implies a $\mathbf{Z}_4$ grading
of the (anti)commutation relations
\begin{equation}\label{comhpq}
\com{\mH_p,\mH_q}\in \mH_{p+q \ (\mathrm{mod \ 4})} \ .
\end{equation}
The generators of  subspaces $\mH^{(i)}$ are denoted as
\begin{eqnarray}
\mH_0: \ T_{[\uc\ud]} \ , \qquad \mH_1: \ T_{\alpha} \ , \qquad
\mH_2: \ T_{\uc} \ , \qquad \mH_3: T_{\halpha} \ .
\end{eqnarray}
Then we can write the current $J_\mu$ as
\begin{eqnarray}
&& J_\mu=J^A_\mu T_A=J^{(0)}_\mu+J^{(1)}_\mu+
J^{(2)}_\mu+J^{(3)}_\mu \ ,
 \\
&& J^{(0)}_\mu=J^{[\uc\ud]}_\mu T_{[\uc\ud]} \ , \qquad
J^{(1)}_\mu=J^{\alpha}_\mu T_\alpha \ , \qquad
J^{(2)}_\mu=J^{\uc}_\mu T_{\uc} \ , \qquad
J^{(3)}_\mu=J^{\halpha}_\mu T_{\halpha} \ ,
\end{eqnarray}
where $A=(a,\alpha,\uc,[\uc\ud])$ and where $J^\alpha_\mu$,
$J^{\halpha}_\mu$ are Grassmann odd vectors, while
$J^{[\uc\ud]}_\mu$, $J^{\uc}_\mu$ are Grassmann even vectors. The
Killing form $\left<\mH_p , \mH_q\right>$, defined in  terms of a
supertrace\footnote{We define the supertrace $\str$ in such a way
that $ \str(M)=\tr A-\tr B $ if $M$ is an even supermatrix and $
\str(M)=\tr A+\tr B $ if $M$ is an odd supermatrix.}, is also
$\mathbf{Z}_4$ invariant and hence we have
\begin{equation}\label{Hpq}
\left<\mH_p,\mH_q\right>=0 \ , \mathrm{unless}
 \  p+q=0 \ \mathrm{mod \ 4} \ .
\end{equation}
 Using the  relation (\ref{Hpq}) we find
that the the Cartan-Killing (super)metric (\ref{Met}) takes the
form
\begin{equation}\label{metG}
K_{AB}=\left(\begin{array}{cccc}
\kappa_{[\uc\ud][\ue\uf]} & 0 & 0 & 0 \\
0 & 0 & 0 & \kappa_{\alpha\hbeta} \\
0 & 0 & \eta_{\uc\ud} & 0 \\
0 & \kappa_{\halpha\beta} & 0 & 0 \\
\end{array}\right) \ .
\end{equation}
Finally we also note that the
 structure constant of the
$psu(2,2|4)$ algebra obey the graded (anti) symmetry property
\begin{equation}\label{id0}
f_{AB}^DK_{DC}= -(-1)^{|A||B|}f_{BA}^DK_{DC}=
-(-1)^{|B||C|}f_{AC}^DK_{DB} \ .
\end{equation}

\section*{Appendix B:
Illustration of the Hamiltonian procedure} \label{AppBfreebos}

In this appendix we will demonstrate  that the canonical approach
given in section \ref{Hamanalys} can be easily applied to the case
of a free massless boson\footnote{We thank H.~Samtleben for e-mail
exchange on this.}. Let us start with the action
\begin{equation}\label{actsim}
S=-\frac{1}{2}\int d^2x \eta^{\mu\nu}
\partial_\mu \phi\partial_\nu \phi \ .
\end{equation}
In the standard Hamiltonian treatment we consider $\phi$ as
canonical variable with the conjugate momentum $P=\partial_0\phi$
and with the standard Poisson brackets
\begin{equation}
\pb{\phi(x),P(y)}=\delta(x-y) \ .
\end{equation}
On the other hand let
us  introduce the group element
$g=e^{\phi}$. Then
\begin{equation}
j_\mu=g^{-1}\partial_\mu g=
\partial_\mu\phi
\end{equation}
and hence the action
(\ref{actsim}) can be written
as
\begin{equation}
S=\frac{1}{2}\int d^2x (
j_0j_0-j_1j_1)
\end{equation}
It is obvious that the current $j_\mu$ obeys the flatness
condition
\begin{equation}
\partial_\mu j_\nu-\partial_\mu j_\nu=0
\end{equation}
that allows one to express $j_0$ as
\begin{equation}\label{j0j1}
j_0=\frac{1}{\partial_1}\partial_0 j_1
\end{equation}
and hence we can interpret $j_1$ as canonical variable. If we
define the conjugate momentum as $\pi={\delta S}/{\delta
\partial_0j}$ and use (\ref{j0j1}) we obtain
\begin{equation}\label{pij1}
\pi=-\frac{1}{\partial_1^2}(\partial_0 j_1)
\ ,
\end{equation}
where it is understood that $j_1$ obeys appropriate boundary
conditions. We define the canonical Poisson bracket according to
\begin{equation}
\pb{j_1(x),\pi(y)}=\delta(x-y) \ .
\end{equation}
Inverting (\ref{pij1}) we obtain
\begin{equation}
-\partial_1\pi=j_0
\end{equation}
and hence
\begin{equation}\label{j0j1h}
\pb{j_0(x),j_1(y)}=
\partial_x \delta(x-y)
 \ .
 \end{equation}
 On the other hand $j_0=\dot{\phi}=P \ ,
 j_1=\partial_1 \phi$ and hence using
 $\pb{\phi(x),P(y)}=\delta(x-y)$
we obtain
\begin{equation}
\pb{j_0(x),j_1(y)}=
\pb{P(x),\partial_y\phi(y)}=
-\partial_y\delta(x-y)=
\partial_x \delta(x-y)
\end{equation}
that coincides with (\ref{j0j1h}). The only subtlety one has to
take into account is the presence of the zero modes of $\phi$ and
$\pi$. Luckily they are finite in number and can be dealt with
separately in connection with the choice of boundary conditions.

\newpage

\end{document}